\documentstyle[times,namedreferences]{kluwer}
\begin{opening}
\title{The Spectrum and Dips of RE 0751+14: 
     A joint evaluation of ROSAT and ASCA Archival Data
 }
\author{\"U. \surname{KIzIlo\u{g}lu}}
\author{A. \surname{Baykal}}
\author{M. \surname{Alev}}
\author{E. \surname{G\"o\u{g}\"u\c{s}}}
\institute{Physics Department, Middle East Technical University, Ankara 06531, Turkey }
\date{}
\end{opening}
\begin{document}

\begin{abstract}

Using archival ASCA and ROSAT
 observations of RE 0751+14, X-ray energy 
 spectra,
 pulse profiles and the results of pulse timing analysis
 are presented. The energy
 spectra are well-fitted by a
 blackbody model at low energy and a Raymond-Smith model at high 
 energy, 
 together with a partial covering absorber. A fluorescence emission
 line at 6.4 keV with an equivalent width $\sim 220 $ eV was
 resolved for the first time.

\end{abstract}

\section{Introduction}

 The intermediate polar (IP) RE 0751+14 (PQ Gem)
  was discovered with the Wide Field
 Camera of the ROSAT satellite (Mason et al., 1992). The detection 
 of coherent 13.9 minute (or 833 sec)
 oscillations in the infrared and optical
 wavelengths are verified by hard X-ray observations of Ginga satellite.
 By using the optical spectroscopy and
 photometry Rosen et al. (1993)
 suggested that the 14.5 minute modulation in the B band
 was the beat of the spin period at 13.9 minute
 with an orbital period of 5.2 hr. In close asynchronously rotating
 binary systems ($P_{orbit}~>~P_{spin}$), a magnetic white dwarf accretes
 material from a low mass companion through Roche lobe overflow, possibly
 with an accretion disk (Patterson 1994).
 Ginga observations (Duck et al., 1994) showed periodic narrow dips at 13.9 
 minutes 
 which are modulated in the  
  energy range of 2-6 keV, however the dip phenomena cease at
 higher energies. This suggests that the absorbing accretion column 
 moves in and out of 
 the line of sight of the observer at the rotation period of the white
 dwarf. Subsequent ROSAT observations have verified the dip modulations
 (Duck et al., 1994).
 These observations have found a soft X-ray component 
 in the light curve of RE 0751+14.
The recent ROSAT observations also  showed distinct soft X-ray components
 in other magnetic polars (Haberl $\%$ Motch 1995).

 In the IPs the inner disk is perturbed by the strong magnetic field
 ($\sim 10^{6}$G) of the white dwarf at a distance close to the magnetospheric
 radius. The plasma flows along the field lines and accretes onto the
 white dwarf near the magnetic pole/s in the form of a thin accretion curtain
 (Singh and Swank 1993). A cross section of the accreting column
 appears to be in the form of a thin semicircular arc. Modulation of the X-ray
 emission can result from variation of the photoelectric absorption
 of the shocked plasma along the line of the sight (Rosen et al., 1988)
 or the occultation of the emission region (King and Shaviv 1984) due to the
 rotation of the white dwarf.
 Recent analysis of pulse phase spectra of RE 01751+14 have shown that
 both models are applicable to RE 0751+14 (Duck et al., 1994).

The pulse arrival times of ROSAT 
and ASCA observatons 
 together with previous X-Ray
and Optical band measurements 
have shown that the 
RE 0751+14 is spining down with rate of 
$\dot P = 1.1\times 10^{-10}$  s/s (Mason 1997). 
In this  
work, therefore we concentrate on the spectral analyses
of this source and 
we examined the unpublished
 archival ROSAT and ASCA data bases
 of RE 0751+14 in combination.
 These observations have longer effective exposure
 times than that of previous ROSAT and GINGA observations and have better
 energy resolution in energy ranges between 0.5-10 keV. We reexamine the
 pulse phase spectra, energy dependent pulses. 
 In the analysis, we noted, for the first time,
  an emission line at 6.4 keV in the
 energy spectrum.
  The power spectrum of the 
intensity time series at low energies (below 0.5 keV)
 showed red noise  which may be an indicator of blob
 accretion proposed for AM Herculis systems (Litchfield and King
 1990).

 \section{Observations}

RE 0751+14 was observed with ROSAT/PSPC and ASCA/SIS and GIS detectors.
The PSPC X-ray observations reported here were obtained between
JD 2449265.70672 and 2449267.11526 with a total effective exposure time of 
25874 s over a time span of 121186 s, whereas ASCA observations took place 
between JD 2449660.92359  and JD 2449663.09701 with a mean  
effective exposure time of 35033  s over a time span of 187769 s.

The ROSAT/PSPC is a gas filled proportional counter sensitive over the
energy range 0.1-2.4 keV with an energy resolution $\Delta$E/E$\sim$0.43
at 0.93 keV.  Detailed descriptions of the satellite,
X-ray mirrors, and detectors can be found in Tr{\"u}mper (1983)
and Pfeffermann et al. (1986).  
The reduction of the ROSAT archival data 
has been performed with the EXSAS package (Zimmermann et al.1993). RE 0751+14 
source counts were extracted from a circle of radius
$2'.5 $ which is expected to include $99\%$ of the photons from the
source, according to the point spread function of the PSPC.
The background was determined from a source free area of radius
$10'$. The mean background subtracted, vignetting and deadtime corrected 
count rate for the whole observation was $4.02\pm0.01$ counts sec$^{-1}$.

The ASCA instrumentation (Tanaka et al. 1994) consists of four imaging 
telescopes, each with a dedicated spectrometer. There are two solid-state
imaging spectrometers (SIS), each consisting of 4 CCD chips, giving 
an energy resolution of 60-120 eV across the 0.4-10 keV band. 
Two gas scintillation proportional counter imaging spectrometers, GIS, 
have an energy resolution of 200-600 eV over the 0.8-10 
keV band. Data were extracted within a $\sim 4$ arcmin radius region 
for each SIS and within a $\sim 6$ arcmin region for each 
GIS. The typical mean count rates for the SIS and GIS are
 $\sim 0.33\pm0.02$ counts sec$^{-1}$ and $\sim 0.31\pm0.02$ counts sec$^{-1}$, 
respectively. 
 Standard cleaning for ASCA data was applied to eliminate X-ray
 contamination from the bright Earth, effects due to high particle background, 
and hot flickering SIS pixels. The reduction of the ASCA archival data  
was performed using XSPEC, XRONOS, XIMAGE and XSELECT softwares. 
In Table 1, we present a journal of the  X-ray observations 
analysed in this work.
\begin{table}
\caption{ Jounal of ROSAT and ASCA observations}
\begin{tabular}{l c c c c c }
\hline
Observation &Detector & JD start & JD end & Time span (sec)& Eff. expo. (sec) \\
\hline
ROSAT & PSPC  & 2448714.01751 & 2448714.96113 & 81532    & 9143 \\
ROSAT & PSPC  & 2449265.70672 & 2449267.11526 & 121186   & 25874 \\
ASCA  & SIS, GIS & 2449661.42359 & 2449663.059701  & 187769  & 35033$^{a}$ \\
\end{tabular}
\vspace{1cm}

~~$^{a}$ Mean exposure time, since ASCA observations were made in several observing modes.
\end{table}

\section{ Results}

\subsection{Spectra}

In this section, 
we will present a detailed spectral analysis of the X-ray data for RE 0751+14
from the observations by ROSAT and ASCA. The large number of photons 
obtained from both 
observations make it possible to test various spectral models as well as 
phase resolved analysis. ROSAT observations have an advantage for low energy
(0.1-2.3 keV) behaviour of the source whereas ASCA detectors extend the 
energy range up to 10 keV (0.5-10 keV). 

RE 0751+14 was detected strongly in the ROSAT/PSPC as a soft X-ray source; 
almost 80$\%$ of the total observed photons have energies below 0.5 keV. 
Single component models   
fail to represent the observed spectra. 
The two component models composed of blackbody plus power law
emission and blackbody plus Raymond Smith fit the data well. 
Due to the low energy resolution of the PSPC, parameters of the blackbody model
and hydrogen column density seems to be highly correlated. 
From these fits, we obtained     
a blackbody temperature kT=34.5$\pm$4.6 eV for soft X-ray component  
and neutral hydrogen column density 
 N$_H=1.4\pm 0.3\times 10 ^{20}$ cm$^{-2}$. 

RE 0751+14 was detected in both SIS and GIS detectors of ASCA. SIS energy 
range starts from 0.4 keV, hence only a small fraction of the soft photons,
expected from the source, contribute to the observed spectra. The energy
ranges of SIS and GIS detectors extend up to $\sim$10 keV allowing us to test
various models. The high energy
resolution of SIS makes it possible to resolve several line features around 
$\sim 6.7$ keV.
ASCA detectors alone however 
 do not reveal the spectral behaviour of the source.
Hence, although there is a long time difference between the observations,
ROSAT and ASCA observations were combined. We assumed that the two data sets 
 differ by a constant normalization factor.
It should also be noted that our preliminary spectral fits showed 
that both observations have similar spectral parameters. 

 The combined ASCA and ROSAT data sets  
represent the data in a broader energy range. 
This allows us to experiment with the spectral data using   
more sophisticated models such as the partial covering absorber model 
(Norton and Watson 1989).
The partial covering absorber model for 
RE 0751+14 first proposed by Duck et al. (1994). 
 In this model,   
an uncovered fraction $X_{1}$ of the source can be seen through a 
interstellar column density $N_{H1}$ and the covered remainder 
of the source, $X_{2}$ such that $X_{1} + X_{2} = 1 $, can be seen through 
intrinsic and interstellar column density, $N_{H2}$.     
In the fits of combined data sets (a) blackbody
plus power-law with a Gaussian line emission profile and (b) blackbody plus
 Raymond Smith model with partial covering absorber 
 and a Gaussian line were tried
(see the discussion section for plausibility of both models). 
The best fit to phase averaged PSPC and SIS data for case (a) gives the 
blackbody
temperature kT=$38.8\pm1.1$ eV, power law index $\alpha=0.87\pm0.01$ and 
 absorption
N$_H=(11.4 \pm 0.6) \times 10^{19}$cm$^{-2}$ with a $\chi^{2}_{\nu} \sim 1.4$.
Case (b) gives similar values for the blackbody temperature and 
absorption.
The plasma temperature for the Raymond Smith model gives high values
of kT=$40.9\pm11.3$ keV. The plasma temperature deduced for 
RE 0751+14 from the observations
of GINGA is $\sim20$ keV (Ishida 1991, Duck et al., 1994). 
 Since the response of the GINGA detectors extened to higher energies  
 up to $\sim 18$ keV, we adopted the 20 keV plasma 
temperature for the pulse phase spectroscopy study. 
Table 2 summarizes
the parameters of the fits. 

\begin{table}
\caption{ The best fitting spectral parameters for the phase averaged data$^{a}$}
\begin{tabular}{l c c c c c c }
\hline
Model & $kT_{bb}$ & $kT_{rs}$ &
                   N$_{H1}$ & N$_{H2}$ & Covering & $\chi^{2}_{\nu} $ \\
    & (eV) & (keV)  &
               ($10^{19}$cm$^{-2}$) & ($10^{22}$cm$^{-2}$) & Frac.(\%) & \\
\hline
BB+RS$^{b}$    &34.5$\pm$4.6 &10.7$\pm$4.9 &13.8$\pm$2.6 & & &1.04 \\
BB+RS+L$^{c,f}$    &38.4$\pm$8.5 &20.0 &11.4$\pm$4.9 &4.9$\pm$0.3 &0.56$\pm$0.08
 &1.33 \\
BB+RS+L$^{d}$    &38.0 &40.9$\pm$11.3 &11.6$\pm$2.3 &6.4$\pm$0.5 &0.47$\pm$0.02
&0.99 \\
BB+RS+L$^{e}$    &38.0 &20.0 &11.4 &5.5$\pm$0.2 &0.54$\pm$0.06 &0.98 \\
\hline
      & $kT_{bb}$ & $\alpha_{pl}$ & N$_{H1}$ &
        $\chi^{2}_{\nu} $ & & \\
\hline
BB+PL+L$^{c}$    &38.9$\pm$1.1 &0.87$\pm$0.01 &11.4$\pm$0.6 &1.4 & & \\
\end{tabular}

\vspace{1cm}

~~$^{a}$Spectral fits were performed using the program XSPEC. Abbreviations for\
\
~~~~~~BB(bb), RS(rs), PL(pl), L stand for Blackbody, Raymond-Smith, Power-law an
d \\
~~~~~~Gaussian line models respectively.\\
~~$^{b,c,d,e}$ are PSPC, PSPC+SIS, PSPC+GIS and PSPC+SIS+GIS detectors respectiv
ely.\\
~~$^{f}$Gaussian Line  energy=(6.38$\pm$0.08) keV, Equivalent Width=(220$\pm$80)
 eV.
\end{table}
\begin{figure}[htb]
\vspace{10cm}
\includegraphics{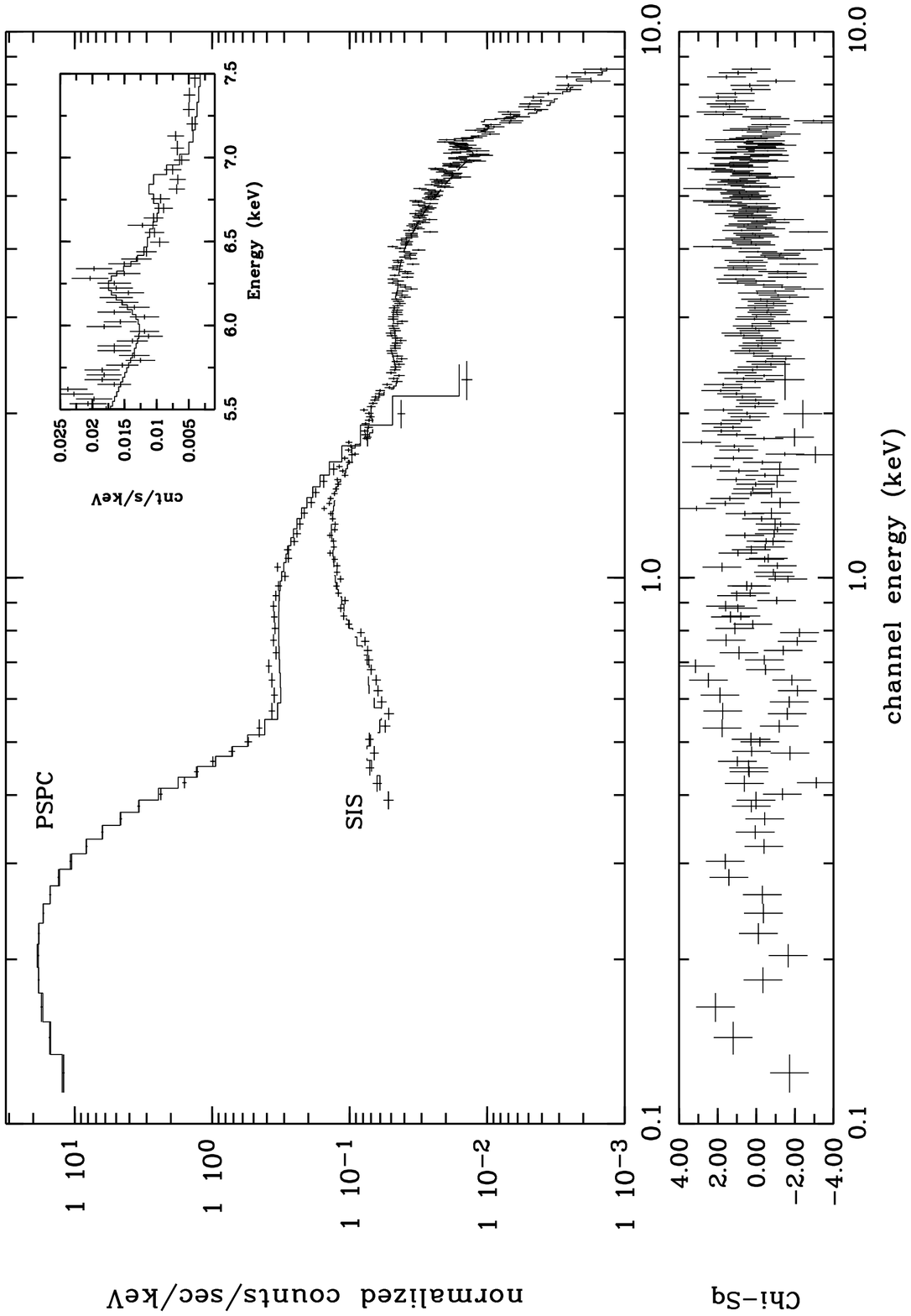}
      \caption{
              The data and blackbody plus Raymond Smith model with partial
          coverage and a Gaussian line model for  PSPC and SIS data.
         Inset shows the 6.4 keV fluorescence line feature.
          }
\end{figure}

In Fig. 1 we present the data and the model for case (b). The inset shows 
the 6.4 keV fluorescence line feature of equivalent width $\sim220$ eV.
A covering fraction of $0.56\pm0.08$ and absorbing column density of
N$_{H2}=4.9\pm0.3 \times 10^{22}$ cm$^{-2}$ are 
obtained from the phase averaged
spectra with a $\chi^{2}_{\nu} \sim 1.3$. The combined 
data set consisting of PSPC, SIS 
and GIS detectors was also used to extend the high energy range and improve
partial coverage parameters. A covering fraction of $0.54\pm0.06$ and absorbing 
column density of N$_{H2}=5.5\pm0.2 \times 10^{22}$cm$^{-2}$ is
obtained with a $\chi^{2}_{\nu} \sim 0.98$.

To see the spectral variation over pulse phase with suitable
statistics, the pulse phase is divided into five phase intervals, 
 one of which is centered on the 
dip observed.  The blackbody temperature and the parameters for the line 
feature were kept constant with the values determined from the phase averaged 
spectra. Significant variation of partial coverage fraction and absorption 
column density around the dip is seen, as presented in Fig. 2. Table 3 summarizes 
the spectral parameters of the pulse phase spectra.

\begin{table}
\caption{ The best fitting spectral parameters from phase resolved spectra }
\begin{tabular}{l c c c c c c c c }
\hline
Phase & $kT_{bb}$ & N$_{H1}$ & N$_{H2}$ & Covering & flux$_{bb}^{a}$ &
        Radius$_{bb}$ & flux$_{tot}^{a}$ & $\chi^{2}_{\nu} $ \\
      &  (eV) & ($10^{19}$cm$^{-2}$) & ($10^{22}$cm$^{-2}$) & Frac.(\%)
      & & ($10^{7}$cm) &  & \\
\hline
0.05-0.25 &32.1$\pm$7.4  &16.9$\pm$0.7 &4.2$\pm$0.7 &0.51$\pm$0.02 &15.1 &1.9 &1
8.5 &1.53 \\
0.25-0.45 &31.5$\pm$11.7 &16.6$\pm$1.0 &4.5$\pm$0.7 &0.47$\pm$0.02 &9.8  &1.4 &1
4.7 &0.91 \\
0.45-0.65 &42.4$\pm$26.1 &9.6$\pm$1.3  &6.7$\pm$0.7 &0.73$\pm$0.01 &4.7  &1.4 &9
.6  &0.92 \\
0.65-0.81 &47.1$\pm$20.9 &8.4$\pm$1.0  &4.5$\pm$0.5 &0.57$\pm$0.02 &3.6  &1.2 &7
.4  &1.02 \\
0.85-0.05 &32.1$\pm$3.2  &17.0$\pm$0.7 &4.6$\pm$0.3 &0.55$\pm$0.01 &3.6  &1.3 &5
.4  &1.06 \\
\hline
\end{tabular}
\vspace{1cm}

$^{a}$Fluxes were calculated from the model spectra in units of $10^{-11}$
 ergs cm$^{-2}$ s$^{-1} $ in the energy range 0.1-10 keV.
\end{table}

\begin{figure}
 \vspace{7.5cm}
\includegraphics{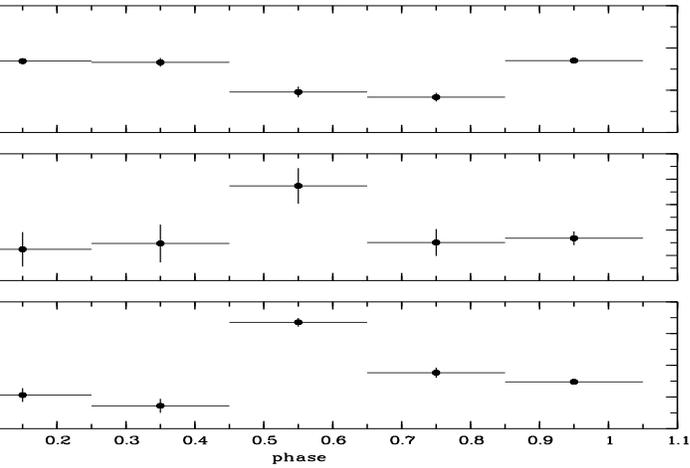}
      \caption{
              Variation of partial coverage fraction and absorption column
           densities for  PSPC and SIS data.
          }
\end{figure}

The total PSPC (0.1-3 keV) flux varies in the range 
(3.9-15.6)$\times10^{-11}
$ergs cm$^{-2}$sec$^{-1}$ with phase. The variation for SIS
(0.25-12 keV) is (2.7-6.0)$\times10^{-11}$ergs cm$^{-2}$sec$^{-1}$. 
The flux for the Raymond Smith component of the partial coverage model is
largest in the phase interval $\phi=0.45-0.65$, which is the region where
the dip is observed, and lowest in $\phi=0.85-0.05$ where the maximum blackbody
flux is observed.
Radius of the effective emitting area for the blackbody model from PSPC data is
(1.2-1.9)$\times10^{7}$ cm for an assumed distance of 400 pc (Patterson 1994).

\subsection{Pulse Profiles}

 General features of the light curves from both ROSAT and ASCA observations 
are quite similar having several dips and maxima on the count rate, 
superposed on 13.9 min sinusoidal modulation. 
 A pronounced dip in the X-ray flux is seen at all energies, just
before the rise to a maximum ($\sim$ 50 degrees before). The depth and the 
width of the dips are energy-dependent (see Fig. 4).

\begin{figure}
 \vspace{8.5cm}
\includegraphics{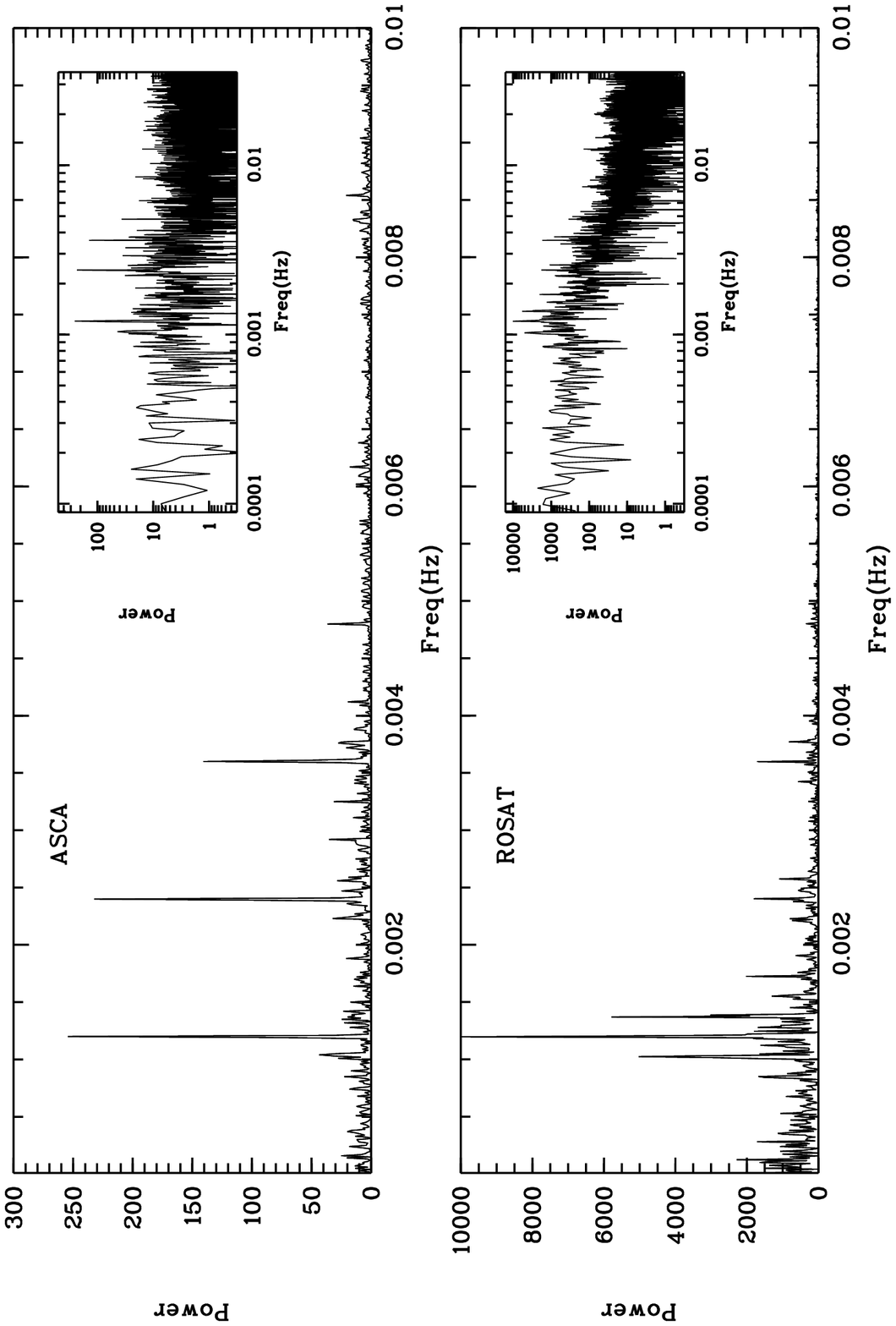}
      \caption{
            The power spectrum of RE 0751+14. The upper panel is the data
            from SIS, binned in 4 sec intervals. $\sim 833$ sec pulsations
            and harmonics are seen clearly. Inset shows the power in log-log
            scale. The lower panel is the same for PSPC data binned in 4 sec
            intervals. The noise at low frequencies is evident  in
            PSPC data in which the contribution of the low energy photons
            is dominant.
          }
\end{figure}
 To see the intensity fluctuations, and
possible signatures of the orbital modulation, pulsations and pulse periods,   
various time series analysis methods are employed for ASCA and ROSAT 
observations of RE 0751+14.

As a first step, the data bases are separated into two energy channels 
according to the soft and hard continuum emission components, then
the discrete power spectra are calculated (Deeming 1975) for the
 low energy range
(0.1-0.5 keV) of ROSAT/PSPC and the 
high energy range (0.5-10 keV) of ASCA as 
shown in Fig. 3a,b .
In the power spectra, we have found only one significant period which is 
the previously identified spin period (13.9 min), with its first 
three harmonics.  The power spectra have shown low frequency noise
in the low energy range. This behaviour is not observed in the 
power spectrum for the high energy range.  
In order to see whether or not the low frequency noise 
is the part of the wobbling of ROSAT/PSPC detectors, the power 
spectrum of ROSAT observations in the range of 
0.5-2.4 keV are obtained 
and it is seen that the power in the low frequencies significanty decreased.   

For the estimation of pulse periods, the time
histories are folded by a number of statistically independent trial periods
(Leahy et al., 1983). Then the pulse profile giving the
maximum $\chi^{2}$ is chosen as the master pulse for each ROSAT or ASCA
observation.
In order to accurately determine the pulse period, a set of 
pulse arrival times was generated by estimating the maximum 
value of cross correlation between the master pulse and averaged 
sample 
pulses. 

\begin{figure}
 \vspace{12cm}
\includegraphics{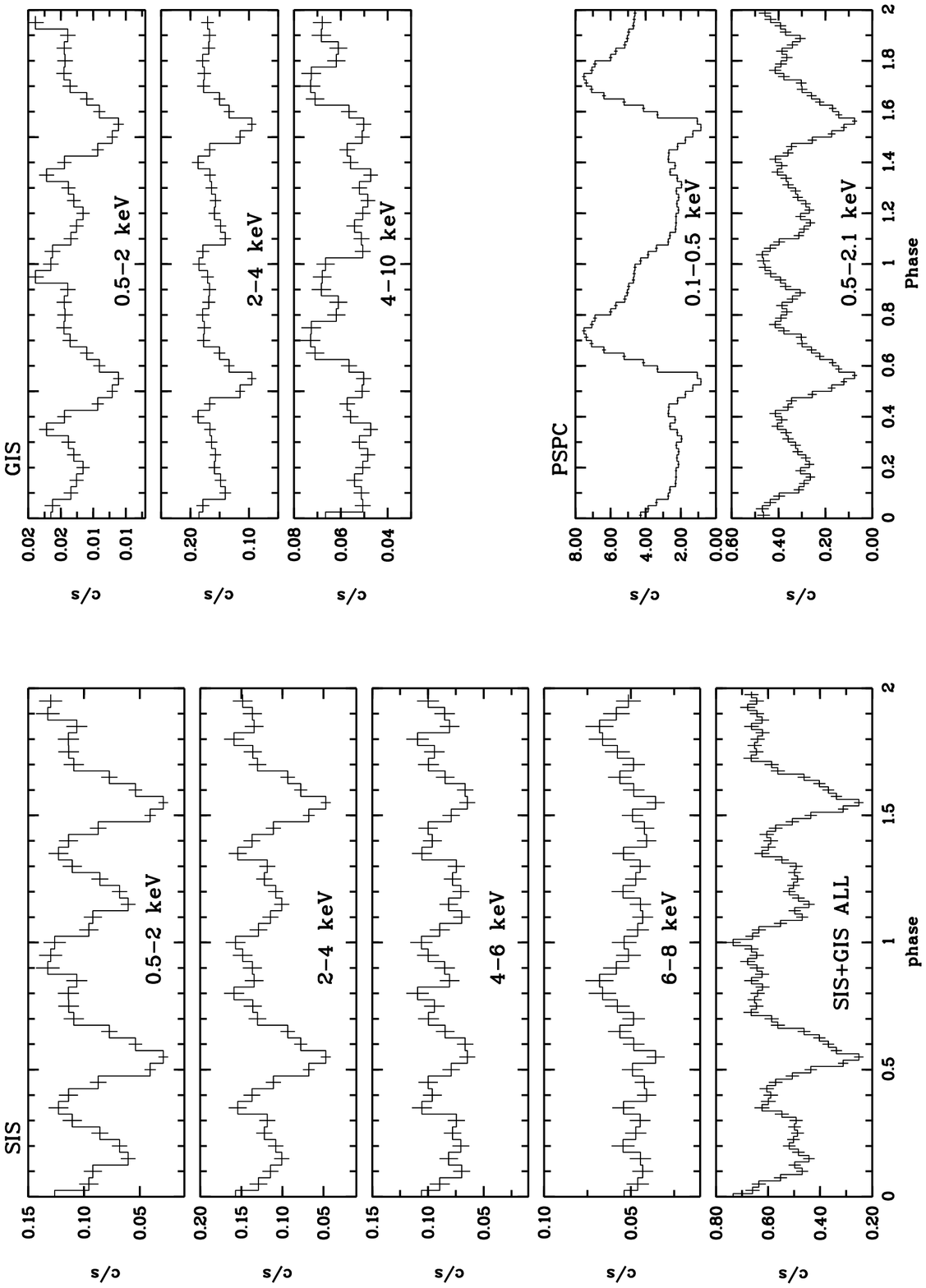}
      \caption{
            PSPC, SIS and GIS energy dependent pulse phases in various
            energy bands. Note the similarity between all the detectors
            for the 0.5-2.0 keV energy band. 0.1-0.5 keV band of PSPC
            is the lowest energy band which is not observable for SIS and GIS.
            The lower pannel on the left is the combined data form SIS and
            GIS detectors of ASCA.
          }
\end{figure}

 In the pulse timing analysis the pulse profiles appear 
to be variable.
In order to control the variability of the pulse profiles we have used the
method of pulse wave filtering as proposed by Deeter and Boynton (1985).
In this method pulse profiles are expressed in terms of harmonic series
and cross correlated with the average pulse profile (master pulse).
The maximum value of the cross-correlation (or pulse arrival times)
 is analytically
well-defined and does not depend on the size of the phase bins of the pulses. 
Short term sharp fluctuations of pulses are naturally filtered by 
a cut off of higher harmonics. The  
pulse periods and their errors are 
estimated from the slopes and uncertainties of pulse arrival times.
The pulse period history of RE 0751+14 is summarized in Table 4.  

\begin{table}
\caption{The pulse period history of RE 0751+14}
\begin{tabular}{l c c }
\hline
Observation & JD(2440000+) & Period (sec)     \\
\hline
GINGA       & 8362.486     & 833.395$\pm$0.01$^{a}$ \\
ROSAT-1     & 8714.018     & 833.33$\pm$0.09$^{b}$  \\
ROSAT-2     & 9265.713     & 833.52$\pm$0.08  \\
ASCA        & 9661.424     & 833.33$\pm$0.10  \\
\hline
\end{tabular}

\vspace{1cm}

$^{a}$ Duck et al., 1994. Based on long data set 19 days apart.\\
~~$^{b}$ Duck et al., 1994 and this work.
\end{table}

 (Note that as an alternative method, we employed the 
$z^{2}$ test (Buccheri et al., 1983) and  verified the above results).

\subsection{Energy Dependence of Pulse Profiles}

In order to see the intensity variations more clearly,
13.9 min pulse profiles are generated by folding the light curves
in various energy ranges.
In the ROSAT observations, more than $\sim 80\%$ of the photons are 
detected below 0.5 keV. Therefore PSPC data is naturally separated into 
two energy windows, namely 'soft' band (0.1-0.5 keV) and 'hard' band 
(0.5-2.1 keV).
For the ASCA observations, the SIS data is divided into four energy bands
(0.5-2, 2-4, 4-6, 6-8 keV), and the GIS data three energy bands, 
(0.5-2, 2-4, 4-10 keV).  In Fig. 4, we present PSPC,
SIS and GIS energy dependent pulse phases together with the combined SIS and
GIS photons in all energies. 

A highly complex pulse shape with a general
 double peaked structure is evident in all three detectors. However, in
 the 0.1-0.5 keV range a single peaked structure with fast rise and slow decay 
is more likely. At energies above 6 keV, again a single peak 
around phases $\phi=0.8-0.9$ is evident.
 The absorption dip around $\phi=0.55$  
 is
clearly seen in PSPC 0.1-0.5 keV band
(Note that the dip epoch was chosen
as $\phi = 0$ in a previous paper by Duck et al., (1994)).
 A strong indication of the dip
can also be seen in 0.5-2 keV bands of PSPC, SIS and GIS. The dip
disappears as the energy increases to higher values. There is an enhancement
around phase $\phi=0.70$ in low energy band of the PSPC. As the energy goes
above 4 keV, the structure in the phase interval $\phi=0.0-0.4$ disappears
but the structures around $\phi=0.6-1.0$ survives. High pulse fraction
at low energies also decreases as the energy increases.

\section{Discussion}
\subsection{Spectra}
The energy spectra is well-fitted to blackbody model at low energies
and Raymond Smith model at high energies with a partial covering
absorption model. In this model the source is heavily occulted 
in same phases of the 13.9 min period by the
absorption of inhomogenous accreting plasma intrinsic to the source 
 (Norton and Watson 1989). 
In general, it is expected that thermal bremsstrahlung
(or Raymond Smith) is the model that fits best to the
 cooling shock regions around magnetic
white dwarfs (Patterson 1994), however it is possible to observe 
power-law continuum emission if inverse comptonization is the 
dominant effect.
 If the spectra have two or more continuum components 
 and observing energy windows
do not extend to the higher energies, the overall trend of the spectrum can be
approximated by a power-law model.

In the case of RE 0751+14, the presence of
 absorption features at lower energies
is clear from the dips of pulse profiles (Fig. 4). 
As the white
dwarf rotates the accreting column occults the source periodically.
 This causes the
absorption at lower energies, while the high energy photons escape
through the accreting column.  The possible confirmation of this feature is 
to see significant increase in the photoelectric absorption
during the dips of the pulses.  Indeed, in the fits with the partial covering 
absorption model, one of the absorbers with a value of 
$N_{H2} \sim 10^{22}$ cm$^{-2}$ showed a significiant increase 
during the dips of the pulses (see Table 3 , Fig. 2). By contrast, 
the single absorber models gave insignificant changes in the
photoelectric absorption during dips.
 It is also very interesting to note that the
equivalent width $\sim 220$ eV of the fluoresence emission line at 6.4 keV 
and the observed column density in the partial covering model
are consistent with empirical relations obtained between the
equivalent widths and column densities in several accretion powered X-ray 
binaries (Inoue 1985).  This empirical correlation 
suggests that the observed fluoresence emission line requires high column 
densites and supports the partial covering model which can 
provide the high column density.  

Another interesting feature is the low frequency noise (or red noise)
in the soft X-ray emission (see Fig. 3a), namely blackbody component of the
X-ray radiation. For an assumed 400 pc source distance (Patterson 1994) 
effective area for the blackbody model from PSPC data 
is only $\sim 2.4 \times 10^{-4} $ of the surface area of typical white dwarf.
This area is comparable with polar cap regions. 
 Low frequency noise suggests the presence of blobs or clumps 
 in accretion. Our observation that low frequency noise is present only in the 
 soft energy channels, which we associate with blackbody emission from near 
 the star's surface, then implies that such blobs are formed in the 
 accretion column near the polar caps. The small cap area would promote
  shocks and instabilities to lead to clump formation.  
 The blackbody radiation near the surface of the white dwarf is then not 
 smooth, following the clumpiness of the accretion. 
 This leads to low frequency noise in the X-ray power spectra. 
  On the other hand thermally cooling
 plasma at the shock regions radiates as optically thin plasma 
 at higher photon energies and the
 high density plasma
 cause the fluoresence emission line at 6.4 keV.

\section{Conclusion}
 
 In this work, we have presented archival ROSAT and ASCA observations 
of interpolar RE 0751+14. We resolve a  
fluorescence line at 6.4 keV with an equivalent width $\sim 220 $ eV 
 for the first time in this source.
The pulse phase spectra of this source showed  
a significant increase in the absorption
at the dips of pulse phase.
The phase resolved spectra can be fitted with 
a single absorber blackbody plus power law model or 
by a partial covering model with a blackbody plus Raymond-Smith. 
Our findings for low frequency noise at low energies and the 
correlation of the line width with 
column density find a natural 
explanation in a two component model
 with partial covering, which we therefore favour. 
In the power spectrum analysis, we have seen a red noise component 
at low energy channels. 
By combining the previously found (and verified in this 
work)      
soft spectral component (Duck et al., 1994) together with red noise 
at low frequencies we concluded that the
 blob accretion to polar caps of the white dwarf 
is quite possible (Litchfield and King 1990). 
A separate component which is associated with the post-shock flow 
near the surface, emits hard X-rays rather smoothly, hence, power specral 
analysis does not show 
a low frequency noise at high energy channels.
Therefore,
our findings strongly suggest that there is an evidence of cold matter
surrounding the plasma with $\sim 10^{22}$cm$^{-2}$.

\section{Acknowledgements}

We thank Dr. Ali Alpar and Dr. Akif Esendemir, for valuable discussions. 
This work is supported by the Scientific and Technical
 Research Councli of Turkey, T{\"U}BITAK, under 
 High Energy Astrophysics Unit.

\end{document}